\documentclass[aps,pra,onecolumn,reprint,floatfix, superscriptaddress]{revtex4-2}

\usepackage{subfigure}
\usepackage{psfrag,graphicx}
\usepackage{pict2e}
\usepackage{dcolumn}
\usepackage{amsmath,amssymb}
\usepackage{bm}
\usepackage{color}
\usepackage{latexsym}
\usepackage{epstopdf}
\usepackage{color}
\usepackage[english]{babel}
\UseRawInputEncoding
\usepackage{amsfonts}
\usepackage{bm}
\usepackage{natbib}
\usepackage{bbold}

\usepackage{enumerate}
\definecolor{myblue}{rgb}{0.2,0.2,0.8}
\definecolor{myzard}{cmyk}{0,0,0.05,0}
\definecolor{mywhite}{rgb}{1,1,1}
\definecolor{mywhite}{rgb}{1,1,1}
\definecolor{myred}{rgb}{1,0.,0.3}
\usepackage[colorlinks=true,citecolor=myblue,linkcolor=myred]{hyperref}

\definecolor{mygrey}{gray}{0.35}
\definecolor{myblue}{rgb}{0.2,0.2,0.8}
\definecolor{myzard}{cmyk}{0,0,0.05,0}
\definecolor{mywhite}{rgb}{1,1,1}
\definecolor{mywhite}{rgb}{1,1,1}
\definecolor{myred}{rgb}{1,0.,0.3}

\def\be{\begin{equation}}
\def\ee{\end{equation}}
\def\ba{\begin{align}}
\def\enda{\end{align}}
\def\bi{\begin{itemize}}
\def\ei{\end{itemize}}

\def\beq{\begin{equation}}
\def\beq{\begin{equation}}
\def\eeq{\end{equation}}

 \newcommand{\ket}[1]{|#1\rangle}

\def\nn{{\textbf n}}
\def\mm{{\textbf m}}

\def\kk{{\textbf k}}
\def\qq{{\textbf q}}

\def\rr{{\textbf r}}

\newcommand{\braket}[2]{\langle #1|#2\rangle}

\newcommand{\mean}[1]{\langle #1\rangle}

\begin{document}

\title{Connecting steady-states of driven-dissipative photonic lattices with spontaneous collective emission phenomena}

\author{A.~Gonz\'alez-Tudela}
\email{a.gonzalez.tudela@csic.es}
\affiliation{Institute of Fundamental Physics IFF-CSIC, Calle Serrano 113b, 28006 Madrid, Spain.}

\begin{abstract}
Recent experimental advances enable the fabrication of photonic lattices in which the light propagates with engineered energy dispersions. When interfaced with quantum emitters, such systems yield strong collective spontaneous emission phenomena, such as perfect sub-radiance, in which the decay into the bath is completely suppressed, forming bound-states-in-the-continuum. Since such photonic lattices are generally lossy, an alternative way of probing them consists in coherently driving them to an steady-state from which photoluminescence can be extracted. Here, we formalize connections between these two seemingly different situations and use that intuition to predict the formation of non-trivial photonic steady-states in one and two dimensions. In particular, we show that subradiant emitter configurations are linked to the emergence of steady-state light-localization in the driven-dissipative setting, in which the light features the same form than the spontaneously formed bound-states-in-the-continuum. Besides, we also find configurations which leads to the opposite behaviour, an anti-localization of light, that is, it distributes over all the system except for the region defined between the driving lasers. These results shed light on the recently reported optically-defined cavities in polaritonic lattices, and can guide further experimental studies.
\end{abstract}

\maketitle

\emph{Introduction.-}
Since the seminal works by Purcell~\cite{purcell46a} and Dicke~\cite{dicke54a}, it is well known that the coupling of emitters to structured photonic baths strongly renormalize their individual and collective emission properties. At the individual level, such structured energy dispersions already give rise to non-Markovian dynamics~\cite{john94a,garmon13a,longui06a,Longhi2006,sanchezburillo17a}, localization of light due to the formation of atom-photon bound~\cite{bykov75a,john90a,kurizki90a,Shi2016,calajo16a} or quasi-bound states~\cite{Gonzalez-Tudela2018,Perczel2020a,Redondo-Yuste2021a,Navarro-Baron2021a}, and directional emission patterns~\cite{Gonzalez-Tudela2017b,Gonzalez-Tudela2017a,galve17a,Yu2019,Gonzalez-Tudela2019a}, among others. When more emitters are considered, the most paradigmatic phenomena is that of super/subradiance~\cite{dicke54a}, in which the lifetime of certain atomic states becomes collectively enhanced or suppressed because of destructive interference. In certain cases, e.g., 1D photonic baths~\cite{ordonez06a,longhi07a,tanaka06a,zhou08a,gonzalezballestero13a,facchi16a,facchi18a}, such suppression can be complete forming what has been labelled as bound-states-in-the-continuum (BIC)~\cite{hsu16a}, in which the light becomes trapped between the emitters.

The improvement in fabrication techniques are enabling already the realization of such structured photonic baths with several platforms ranging from photonic-crystal waveguides~\cite{goban13a,goban15a,lodahl15a,Chang2018}, to photonic lattices based on coupled microwave resonators~\cite{houck12a,liu17a,Schmidt2013CircuitCircuits,Mirhosseini2018a,Kollar2019HyperbolicElectrodynamics,Blais2020QuantumElectrodynamics} or semiconductor microcavities~\cite{Deng2010Exciton-polaritonCondensation,Deveaud2007TheDevices,Kavokin2008Microcavities,Boulier2020}, among others. This has enabled, for example, engineering photon energy dispersions with topological bands~\cite{jean17a,Kim2020b} or Dirac-like behaviour~\cite{Milicevic2019a,Real2020}. However, since such photonic lattices are inherently lossy, the conventional way of probing them is not by coupling them to excited emitters, but rather by driving them with lasers into an steady state that can be monitored through photoluminescence. A recent experiment with polaritonic lattices~\cite{Jamadi2021} has shown that judiciously choosing the drives, such photonic steady-states can localize light between several local drives, similarly to what occurs in the subradiant spontaneous emission configuration. Thus, a timely question is to understand whether these phenomena are connected and, if positive, what intuition can we gain from it.

\begin{figure}[tb]
    \centering
    \includegraphics[width=\linewidth]{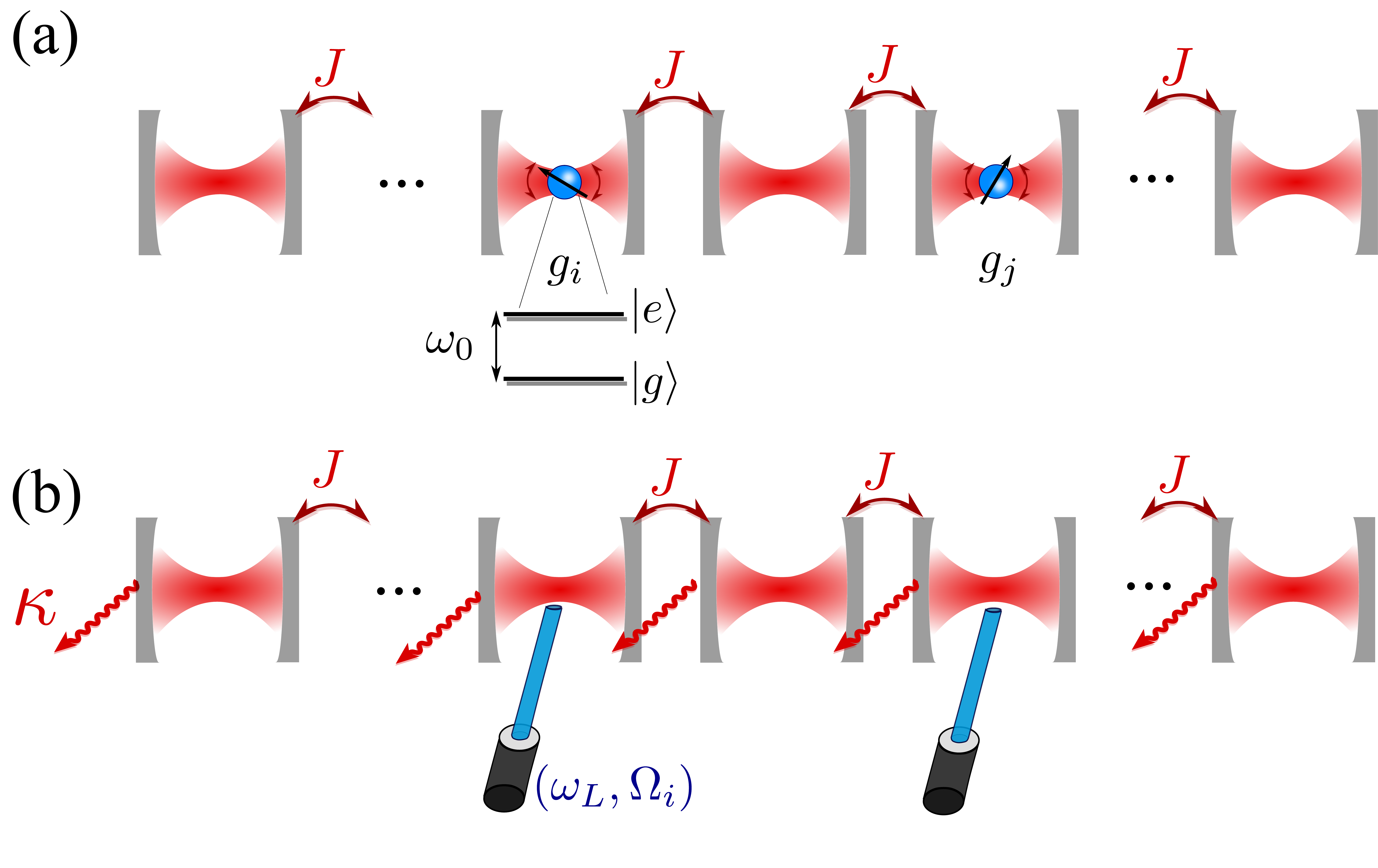}
    \caption{(a) Quantum emitter configuration: quantum emitters with an optical transition of frequency $\omega_0$ couple locally with strength $g_i$ to a photonic lattice described by a coupled-cavity array, with cavity frequency $\omega_a$ and nearest-neighbour hopping $J$. (b) Driven-dissipative configuration: one or several lasers with frequency $\omega_L$ and amplitude $\Omega_i$ drive locally the cavities of the coupled cavity array, which now we assume have individual, equal losses with rate $\kappa$.}
    \label{fig:0}
\end{figure}

In this work we address this question by establishing formal connections between the two situations, that are, the spontaneous emission dynamics of super/subradiant states and the steady-states of driven-dissipative photonic lattices (see Fig.~\ref{fig:0} for an schematic picture of the two situations in the case that the discrete photonic lattice is described by a coupled cavity array). In particular, we are able to show that one can extract the steady-state localization conditions from the analysis of subradiance in the spontaneous emission configuration. In fact, one can prove that the steady-state localization displays the same functional form than the photonic component of the spontaneously formed BIC, shedding light on the origin of the recently reported optically-defined cavities~\cite{Jamadi2021}. On top of that, we also find driving configurations which lead to light "anti-localization"t, that is, the light spreads in the steady-state along the whole system, except in the space between the drives. We illustrate these phenomena with both one and two-dimensional photonic lattices, when possible, and study how the non-trivial steady-states emerge in the time domain.

\emph{Collective spontaneous emission.-} Let us first review what occurs in the quantum emitter configuration depicted in Fig.~\ref{fig:0}(a). The total Hamiltonian describing the dynamics in that configuration reads: $H=H_S+H_B+H_I$. Here,  $H_B=\sum_\kk \omega(\kk) a_\kk^\dagger a_\kk$ is the photonic lattice Hamiltonian (we take $\hbar=1$ throughout the manuscript), that we assume to be traslationally invariant such that momentum $\kk$ is conserved, and that it can be described by a single Bravais lattice with eigenoperators given by $ a^{(\dagger)}_\kk=\frac{1}{\sqrt{N}}\sum_{\rr} a^{(\dagger)}_\rr e^{-(+)i\kk\cdot\rr}$, written in terms of the creation annihilation (creation) operators in real space, $a^{(\dagger)}_\rr$, being $N$ the total number of sites. Note also here that $\kk$ is a vector whose components are discrete in finite systems, running from $k_i\in [-\pi,\pi]$ in $2\pi/N$ steps. $H_S=\omega_0\sum_{j=1}^{N_e}\sigma^j_{ee}$ is the emitter's Hamiltonian, which we assume them to have a single optical transition of frequency $\omega_0$, that we take equal for all emitters. Note that we use $\sigma_{\alpha\beta}^j$ to denote the operators of the $j$-th emitter. Finally, $H_I$ is the light-matter Hamiltonian which reads:
\begin{align}\label{eq:HI}
H_I&=\sum_{j=1}^{N_e} \left(g_j a^\dagger_{\rr_j}\sigma^j_{ge}+\mathrm{h.c.}\right)=\sum_{\kk}\sum_{j=1}^{N_e} \left(g_{\kk,j} a^\dagger_{\kk}\sigma^j_{ge}+\mathrm{h.c.}\right)\,,
\end{align}
where $g_j$ is the coupling of the $j$-th emitter to the bath site at position $\rr_j$, and $g_{\kk,j}=\frac{g_j}{\sqrt{N}} e^{-i\kk\cdot\rr_j}$ its $\kk$-dependent coupling strength. 

One way of finding the perfect subradiant (or BIC) states in this configuration, that are states that does not decay into the bath, is by noticing that they are eigenstates of the full light-matter Hamiltonian, i.e., $H\ket{\Psi_\mathrm{BIC}}=E_\mathrm{BIC}\ket{\Psi_\mathrm{BIC}}$, with energies  $E_\mathrm{BIC}\in\omega(\kk)$. Since we are interested in finding analogies with the coherently driven-dissipative scenario, which is by construction in the linear regime, it is enough to restrict to single-excitation solutions. Then, $\ket{\Psi_\mathrm{BIC}}$ can be written as:
\begin{align}
\ket{\Psi_\mathrm{BIC}}= \left(\sum_jC_j\sigma^j_{eg}+\sum_\rr\alpha_\rr a_\rr^\dagger \right) \ket{\mathrm{vac}}\,.
\end{align}

Imposing that form, one arrives to a set of coupled equations, from which one can obtain that the photonic component of the BIC reads:
\begin{align}
    \alpha_{\rr,\mathrm{BIC}}(E_\mathrm{BIC})=\frac{1}{\sqrt{N}}\sum_\qq \frac{\sum_{j} g_j C_j e^{i\qq\cdot(\rr-\rr_j)}}{E_\mathrm{BIC}-\omega(\qq)}~\label{eq:photBIC}\,.
\end{align}

Here, $E_\mathrm{BIC}$ is the exact energy of the BIC, which can be obtained by solving the following matricial equation: 
\begin{align}
\left[\left(E_\mathrm{BIC}-\omega_0\right)\mathbf{I}_{N_e}-\bar{\Sigma}(E_\mathrm{BIC})\right]\mathbf{C}=0\,,\label{eq:poleeqmat}
\end{align}
where $\mathbf{I}_{N_e}$ is the identity matrix of size $N_e$, $\mathbf{C}=(C_1,\dots,C_{N_e})$ is the vector with the emitters' probability amplitudes, and $\bar{\Sigma}(E)$ is a matrix whose entries are the individual and collective self-energies $\Sigma_{ij}(E)$, which read:
\begin{align}
    \Sigma_{ij}(z)=\frac{1}{N}\sum_{\qq} \frac{g_i^* g_j e^{i \qq(\rr_i-\rr_j)}}{z-\omega(\qq)}=\Sigma_{ji}^*(z)\,.~\label{eq:self}
\end{align}

Interestingly, these self-energies also govern the emitters' dynamics when they are initially in an excited state. Assuming one can adiabatically eliminate the photonic bath under the Born-Markov conditions, such dynamics is governed by an effective (non-Hermitian) Hamiltonian $H_\mathrm{eff}$ which reads: 
\begin{align}
H_\mathrm{eff}=\sum_{i,j}\Sigma_{ij}(\omega_0+i0^+)\sigma_{eg}^i\sigma_{ge}^j\,,\label{eq:Heff}
\end{align}

Thus, a perfect subradiant state will be an eigenstate of $H_\mathrm{eff}\ket{\Psi_{\mathrm{sub}}}=z_\mathrm{sub}\ket{\Psi_{\mathrm{sub}}}$, with $\mathrm{Im}(z_\mathrm{sub})=0$. Since $\mathrm{Im}\left[\Sigma_{ii}(\omega_0+i0^+)\right]\neq 0$ when $\omega_0\in \omega(\kk)$, this implies that one requires more than a single emitter to cancel its imaginary part, and thus, be able to find a BIC.

Summing up, what occurs when one initializes the emitter's system in a state, $\ket{\Psi_0}$, with some overlap with the subradiant state is that it evolves in the infinite-time limit to:
\begin{align}
\ket{\Psi(t\rightarrow\infty)}=\lim_{t\rightarrow \infty}e^{-i H_\mathrm{eff}t}\ket{\Psi_0}=\braket{\Psi_{\mathrm{sub}}}{\Psi_0}\ket{\Psi_{\mathrm{sub}}}\,.
\end{align}

Besides, this is also accompanied by an steady-state value in the photonic bath with the form given by Eq.~\eqref{eq:photBIC}:
\begin{align}
\alpha_\rr(t\rightarrow\infty)=\alpha_{\rr,\mathrm{BIC}}(\omega_0)\,,\label{eq:st}
\end{align}
localized between the emitters. Note that if an additional decay channel is included into the emitters, with rate $\Gamma^*$, this will have a two-fold impact: first, it will provide a finite lifetime to the BIC/subradiant state (it will not survive in the infinite-time limit); and second, it will alter its (transient) shape by shifting its resonance, i.e., $\alpha_{\rr,\mathrm{BIC}}\left(\omega_0-i\Gamma^*/2\right)$.

\begin{figure}[tb]
    \centering
    \includegraphics[width=0.85\linewidth]{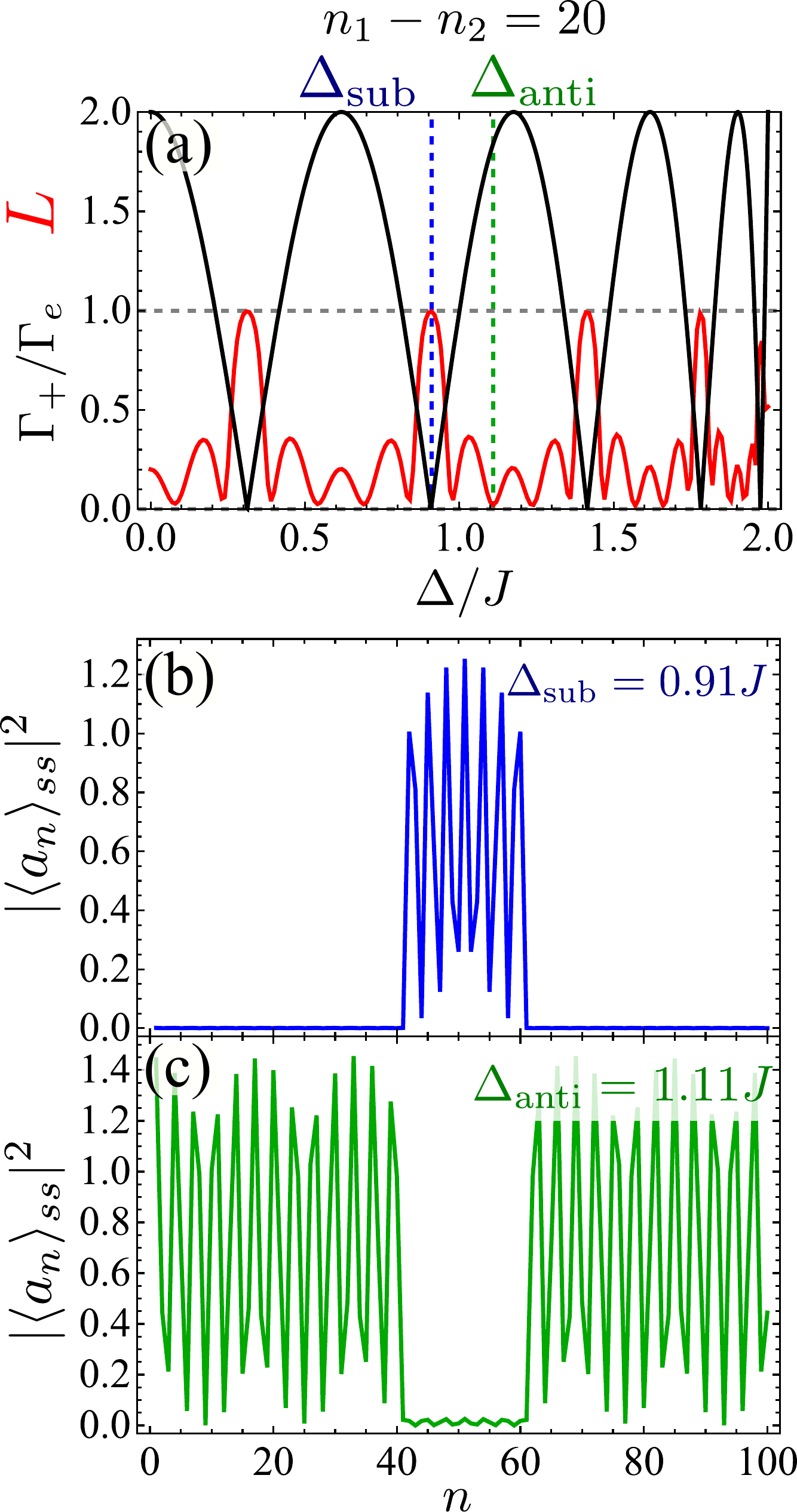}
    \caption{(a) Collective decay rates (solid black), $\Gamma_{+}/\Gamma_e=1+\cos(q_0(\Delta)n_{12})$ for the symmetric subspace as a function of $\Delta/J$ for a fixed inter-emitter distance $n_{12}=20$. In solid red, we plot the localization parameter $L$ defined in Eq.~\eqref{eq:locR}, which measures the amount of steady-state population between the local drives. The vertical lines correspond to the two detunings $\Delta_{\mathrm{sub} [\mathrm{anti}]}$ that we choose to plot the steady-state configurations of panels (b) [(c)], respectively. (b-c) Steady-state population in the symmetric driving case $\Omega_{n_1}=\Omega_{n_2}=J$ for the detunings $\Delta_{\mathrm{sub} [\mathrm{anti}]}$, which leads to [anti-]localization, respectively. In all situations we fix $\kappa=0.01J$ and $N=100$.}
    \label{fig:1}
\end{figure}

Let us characterize the emergence of this perfect subradiant phenomena particularizing for $N_e=2$ emitters, with no losses ($\Gamma^*=0$), and a one-dimensional coupled cavity array as plotted in Fig.~\ref{fig:0}(a). In that case, the photonic energy dispersion reads $\omega(k)=\omega_a-2J\cos(k)$ (we take the lattice constant as a unit of length), which allows one to extract an analytical expression of the (Markovian) self-energies in the limit $N\rightarrow \infty$~\cite{lekien05a,chang12a,Gonzalez-Tudela2013b,calajo16a}:
\begin{align}
    \Sigma_{ij}(\omega_0+i0^+)&=-i\frac{|g|^2}{v_g\left(q_0\left(\Delta\right)\right)}e^{i q_0\left(\Delta\right)n_{ij}}=-i\Gamma_e e^{i q_0\left(\Delta\right)n_{ij}}\, \,,~\label{eq:self1d}
\end{align}
where $\Delta=\omega_a-\omega_0$ is the detuning between the cavity and emitter's frequency, $n_{ij}=n_i-n_j$ the inter-emitter distance, $v_g(k)=\partial_k\omega(k)$ the group velocity, $\Gamma_e$ the individual decay rate, and $q_0\left(\Delta\right)$ is the momentum resonant with the emitter's optical transition, which is given by $q_0\left(\Delta\right)=\pm \arccos{\left(\frac{\Delta}{2J}\right)}$.  In that case, it is convenient to rewrite Eq.~\eqref{eq:poleeqmat} or the effective Hamiltonian of Eq.~\eqref{eq:Heff} in the symmetric/antisymmetric subspaces, i.e., $\ket{\Psi_{0,\pm}}=(\ket{eg}\pm\ket{ge})/\sqrt{2}$ since they decouple and their dynamics can be calculated independently. In particular, one can calculate the decay rate (imaginary part of the collective self-energy) of each subpsace, obtaining:
\begin{equation}
   \Gamma_{\pm}=\Gamma_e\left[1\pm \cos\left(q_0\left(\Delta\right) n_{12}\right)\right]\,,
\end{equation}

From that equation, one can immediately see that the emergence of perfect subradiance, $\Gamma_{\pm}=0$, depends on a trade-off between the symmetric/antisymmetric nature of the excitations ($\pm$), the distance between emitters ($n_{12}$), and the energy of the optical transition ($\Delta$). This is illustrated in Fig.~\ref{fig:1}(a) where we plot the decay rate in the symmetric subspaces $\Gamma_{+}/\Gamma_e$ (solid black) as a function of $\Delta$ for a fixed inter-emitter distance $n_{12}=20$. There, we see how there are several instances of $\Delta$ which makes $\Gamma_{+}=0$, corresponding to the values when $\cos\left(q_0\left(\Delta\right) n_{12}\right)= - 1$. As shown in the literature~\cite{ordonez06a,longhi07a,tanaka06a,zhou08a,gonzalezballestero13a,facchi16a,facchi18a}, the perfect subradiant states are also accompanied by a localization of light between emitters. 
In what follows, we see how the intuition obtained from this analysis can be translated to the driven-dissipative scenario to obtain non-trivial steady-states.

\emph{Driven-dissipative photonic lattices.-} Let us now consider that we have no excited emitters, but rather drive the photonic lattices with lasers, as depicted in Fig.~\ref{fig:0}(b). We will assume that we can control independently the amplitudes and phases of the lasers in each cavity, $\Omega_i$, as can be done in experiments~\cite{Jamadi2021}, but for simplicity restrict to the case where all have the same frequencies, $\omega_L$. Then, the Hamiltonian describing the dynamics within the photonic lattice reads:
\begin{align}
    H_C=H_B+\sum_i \left(\Omega^*_i a_i+\mathrm{H.c.}\right)\,,
\end{align}
that we implicitly write in a rotating frame at the laser frequency, $\omega_L$, such that one must replace $\omega_a\rightarrow \Delta_L=\omega_a-\omega_L$ in $H_B$. Apart from this Hamiltonian, we need to include the fact that the cavities are lossy, and thus decay with a rate $\kappa$ which we assume to be the same for all cavities. The photonic lattice dynamics is then given by the following master equation:
\begin{align}
\dot{\rho}=i[\rho,H_C]+\sum_{i}\kappa\left(a_i\rho a_i^\dagger -\frac{1}{2}\{\rho,a_{i}^\dagger a_i\}\right)\,,
\end{align}

Defining $\mathbf{X}=(a_1,a_2,\dots,a_{N_c})^T$, and $\mathbf{\Omega}=(\Omega_1 ,\dots,\Omega_{N_c})^T$ it is easy to see that both the Hamiltonian:
\begin{align}
 H_C=\mathbf{X}^\dagger h_B \mathbf{X}+\left(\mathbf{X}^\dagger\mathbf{\Omega}+\mathrm{H.c}\right)\,,
\end{align}
and Lindblad parts are quadratic in terms of the cavity operators. Thus, the dynamics of the coherences mean values, $\mean{\mathbf{X}(t)}$, are given by the following equation (sometimes referred to as temporal coupled-mode theory~\cite{Haus1984WavesOptoelectronics,Fan2003TemporalResonators,Joannopoulos2011}):
\begin{align}
 \frac{d\mean{\mathbf{X}(t)}}{dt}=W\mean{\mathbf{X}(t)}-i \mathbf{\Omega}\,
\end{align}
where $W=-i\left(h_B-i\frac{\kappa}{2}\mathbf{I}_{N}\right)$. This equation can then be formally solved as follows:
\begin{align}
\mean{\mathbf{X}(t)}=e^{W t}\left(\mean{\mathbf{X}(0)}-\mean{\mathbf{X}}_{ss}\right)+\mean{\mathbf{X}}_{ss}\,.~\label{eq:dyn}
\end{align}
where $\mean{\mathbf{X}(0)}$ is given by the initial state of the cavities, that we will assume to be empty, and $\mean{\mathbf{X}}_{ss}$ is the steady-state value obtained from: 
\begin{align}
 \mean{\mathbf{X}}_{ss}=i W^{-1}\mathbf{\Omega}\,. ~\label{eq:steady1}
\end{align}

Note in this case the cavity populations are simply given by $\mean{a_i^\dagger a_i}=|\mean{a_i}|^2$, so that calculating the coherence vector $\mean{\bf{X}}$ is enough to characterize the steady-state of our system.

To get further insight on the steady-state solution of Eq.~\eqref{eq:steady1}, we can write $W^{-1}$ using the eigenvalues and eigenvectors of $h_B=U \mathrm{diag}\left(\tilde{\mathbf{h}}_B\right) U^\dagger$, with $\tilde{\mathbf{h}}_B$ being the vector with the eigenvalues of $h_B$ (note $h_B$ is the only non-diagonal part of $W$). Using that, the steady-state solution can be written as:
\begin{align}
 \mean{\mathbf{X}}_{ss}= U \mathrm{diag}\left(\frac{1}{-\tilde{\mathbf{h}}_B+i\kappa/2}\right)             U^\dagger\mathbf{\Omega}\,.~\label{eq:drivendiscoh}
\end{align}

When $H_B$ is traslationally invariant, which is the case we will consider for this work, then $U$ is a unitary matrix given by the Bloch phases~\cite{Bloch1929UberKristallgittern,Ashcroft1976DeterminationDiffraction}, and $\tilde{\mathbf{h}}_B=\omega(\kk)-\omega_L$. With that insight, one can show that
\begin{align}
\mean{a_\nn}_{ss}=\frac{1}{N}\sum_\qq \frac{\sum_\mm \Omega_\mm e^{i\qq(\nn-\mm)}}{\omega_L-\omega(\qq)+i\kappa/2}\label{eq:steady2}
\end{align}

Comparing Eqs.~\eqref{eq:steady2} and~\eqref{eq:photBIC}, one notices that:
\begin{align}
\lim_{\kappa\rightarrow 0}\mean{a_\nn}_{ss}=\alpha_{\nn,\mathrm{BIC}}(\omega_L)~\label{eq:conec}
\end{align}
where the role of the coupling/phases of the emitters ($g_j C_j$) are played by the laser amplitudes ($\Omega_m$), and the role of the  BIC/emitters energy ($E_\mathrm{BIC}=\omega_0$) is played by the laser frequency ($\omega_L$). Thus, whenever a BIC emerges with emitters, a steady-state localization of light can be found in the driven-dissipative scenario. Note also that $\mean{a_\nn}_{ss}^2$ with $\kappa\neq 0$, coincides with that of the transient BIC with emitters when including emitter losses $\Gamma^*$ with the same strength $\Gamma^*=\kappa$. Note though that $\Gamma^*$ and $\kappa$ enter with a different sign in the expressions of Eq.~\eqref{eq:conec} and Eq.~\eqref{eq:st}.

Let us numerically illustrate this connection with the 1D model we used in the spontaneous emission configuration, considering the two local probes with equal amplitudes $\Omega_{n_1}=\Omega_{n_2}=\Omega$ at the emitters' positions $n_1$ and $n_2$ with $n_1-n_2=20$. To characterize the emergence of steady-state localization of light, we define a parameter, $L$, that gives the (normalized) steady-state population between the two local drives as compared to the rest of the chain:
\begin{align}
L=\frac{\sum_{n\in(n_1,n_2)}\mean{a_n}^2}{\sum_n\mean{a_n}^2}\,.\label{eq:locR}
\end{align}

\begin{figure}[tb]
    \centering
    \includegraphics[width=\linewidth]{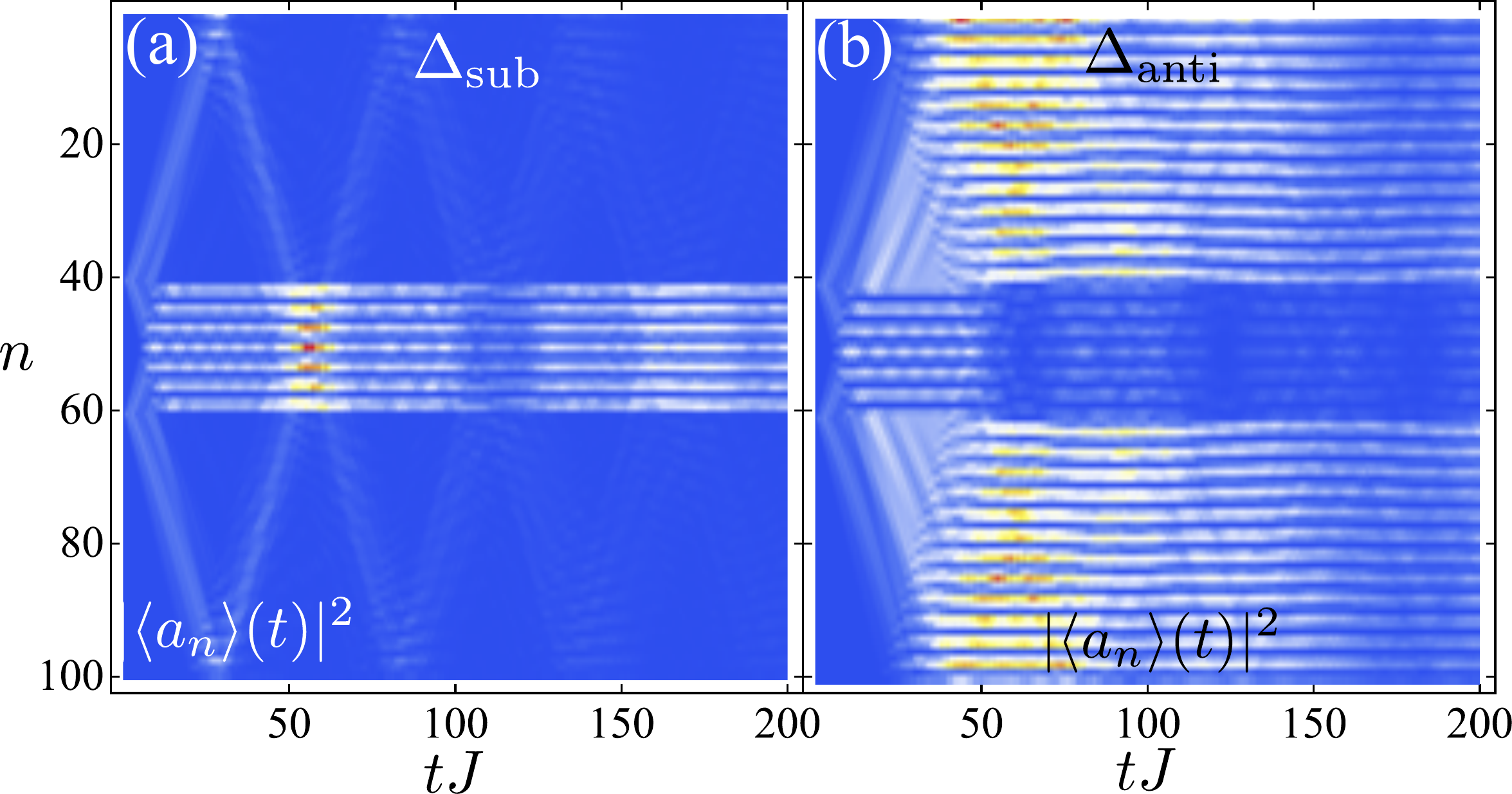}
    \caption{(a) [(b)] Population dynamics for a  symmetric driving configuration with $n_{12}=20$, and detunings $\Delta_\mathrm{sub}$ [$\Delta_\mathrm{anti}$] (see Fig.~\ref{fig:1}). The chain has $\kappa=0.01J$ and $N=100$ lattice sites. One observes the emergence of the optically defined localized [and anti-localized] modes plotted in Figs.~\ref{fig:1}(b-c), respectively.}
    \label{fig:2}
\end{figure}

In Fig.~\ref{fig:1}(a), we plot $L$ (in solid red), together with $\Gamma_{+}/\Gamma_e$, observing that whenever a $\Gamma_+(\Delta_\alpha)=0$ for the emitter's symmetric subspace, a maximum in $L(\Delta_\alpha)\approx 1$ appears, as expected from the connection of Eq.~\eqref{eq:conec}. In Fig.~\ref{fig:1}(b), we plot the steady-state population for one of these detuning parameters, $\Delta_\mathrm{sub}$, highlighted in blue Fig.~\ref{fig:1}(a), where one sees explicitly how indeed the light is localized between the local drives. In Fig.~\ref{fig:2}(a), we plot the dynamical formation of these optically defined cavities, using Eq.~\eqref{eq:dyn}, to observe how the localization occurs because of the destructive interference between the emission of the two local drives, reminiscent of what occurs with entangled emitters.

Apart from this connection betwen the spontaneously formed BIC and steady-state localization, we find another interesting behaviour of the localization parameter $L$ in Fig.~\ref{fig:1}(a). In particular, we observe how certain detunings $\Delta_\beta$'s lead to $L(\Delta_\beta)\approx 0$, that is, the steady-state features almost no population between the local drives. We label this phenomenon as anti-localization, and it is illustrated more clearly in Fig.~\ref{fig:1}(c) where we plot the steady-state population along the chain for a particular detuning $\Delta_\mathrm{anti}$ (highlighted in green in Fig.~\ref{fig:1}(a)) which leads to that behaviour. In Fig.~\ref{fig:2}(b), we also plot the dynamical formation of such behaviour and note an important difference with the steady-state localization, that is, that the interference that leads to the anti-localization occurs after the emission of the local drives travel through all the system. Thus, the detuning $\Delta_\beta$ where such anti-localization appears depends on both the size and boundary conditions of the system. One can understand such anti-localization as a localization at the outer region. Thus, for the case with periodic boundary conditions, it appears for detunings such that $e^{i q(\Delta_\beta) (N-d)}=\pm 1$ since $N-d$ is the distance connecting the probes from the outer region.

\emph{Arbitrary steady-state shapes in two-dimensions.-} Let us now show how the steady-state localization can also be found in two-dimensional systems, and can lead to very versatile control of their shapes. In general, finding perfect subradiance in two-dimensions with local probes is challenging because light propagates in many directions. However, in photonic lattices with energy dispersions with Van-Hove singularities and straight isofrequencies, spontaneous emission can be highly directional~\cite{Gonzalez-Tudela2017b,Gonzalez-Tudela2017a,galve17a,Yu2019,Gonzalez-Tudela2019a}. For example, in the case of square coupled-cavity arrays, with energy dispersion given by $\omega(\kk)=\omega_a-2J\cos(k_x)-2J\cos(k_y)$, the emitters resonant with the middle of the band decay into the bath in a directional way. Such directional emission was shown to lead to the emergence of a BIC with four emitters~\cite{Gonzalez-Tudela2017b,Gonzalez-Tudela2017a} if they are disposed at positions $\nn_{1/2}=(\pm n,0)$, $\nn_{3/4}=(0,\pm n)$ with local phases $C_1=C_2=\pm C_3=\pm C_4$, depending on whether $n$ is odd/even. By virtue of the connection we find in Eqs.~\eqref{eq:photBIC},~\eqref{eq:steady2},~\eqref{eq:conec}, it is expected that steady-state localization can also be found in the two-dimensional driven-dissipative scenario.

\begin{figure}[tb]
    \centering
    \includegraphics[width=\linewidth]{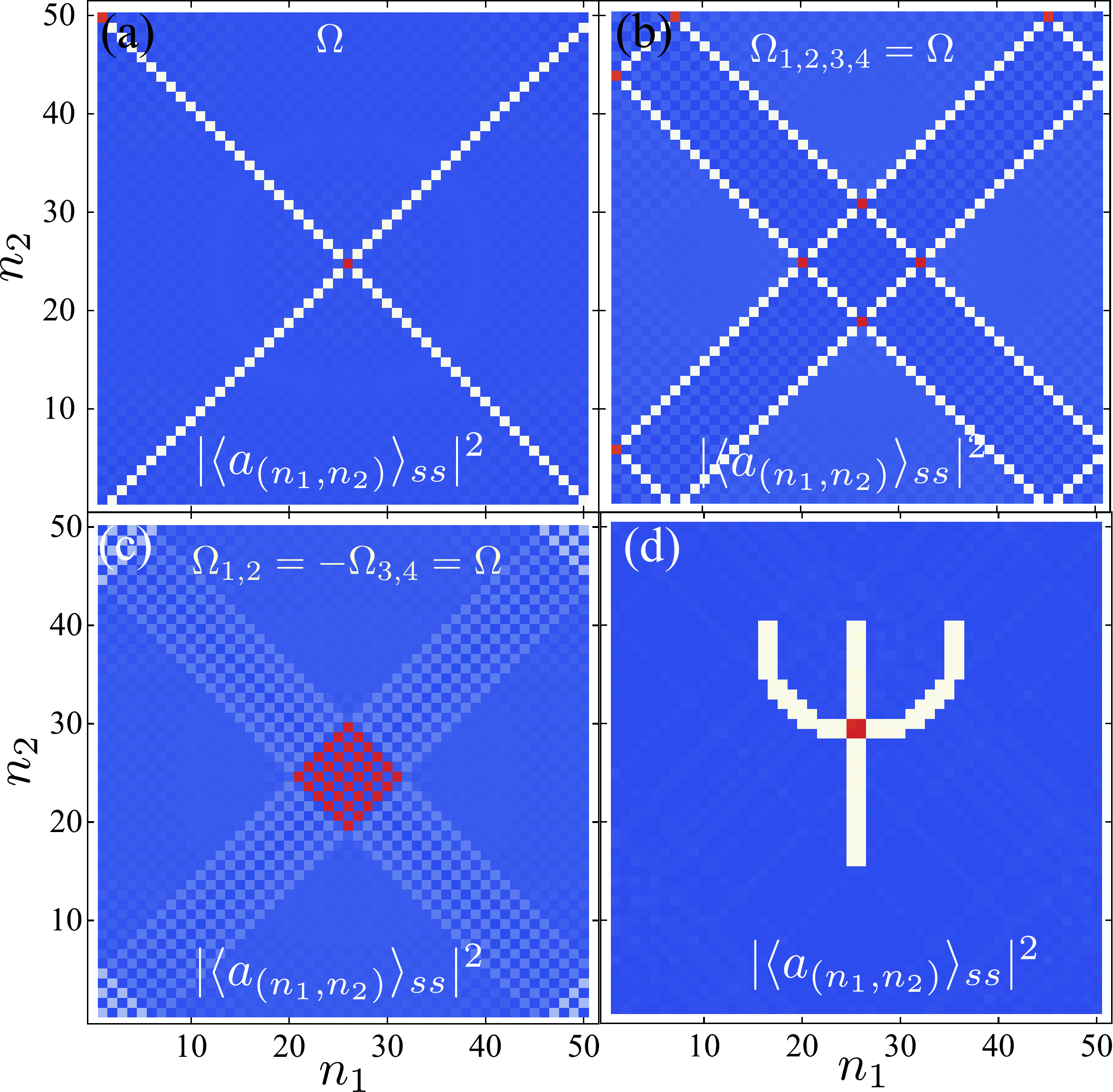}
    \caption{(a-d) Steady-state population of a driven-dissipative photonic lattice with $50\times 50$ sites with different driving configurations. (a) Local driving at the central point chain, where the light spreads over four directions. (b-c) Four local drives at positions separated $\nn_{1/2}=(\pm 6,0)$, $\nn_{3/4}=(0,\pm 6)$ from the central point with symmetric/antisymmetric drives, respectively. For such distance, the antisymmetric configurations displays a perfect localization of light between the lasers, as expected from the analogy with BIC formation with emitters~\cite{Gonzalez-Tudela2017b,Gonzalez-Tudela2017a}. (d) $\Psi$-shape steady-state localization found by driving with multiple non-local drives (see text for explanation). In all cases the absolute value of the drive amplitude $|\Omega|=J$, the cavity losses are set to $\kappa=0.01J$, and the drives are tuned to the middle of the band $\omega_L=\omega_a$}
    \label{fig:3}
\end{figure}

In Fig.~\ref{fig:3} we illustrate this phenomenon for a 2D square photonic lattice with $50\times 50$ sites and local drives resonant to the middle of the band, i.e., $\omega_L=\omega_a$. First, in Fig.~\ref{fig:3}(a) we plot the resulting steady-state populations for a single laser drive at the center. In that case, the light spreads prominently in four directions due to the non-uniform group velocity, which for this model reads $v_g(\kk)=2J\left(\sin(k_x),\sin(k_y)\right)$, along the resonant frequencies. This leads to $\kk$-points with zero-group velocity, such as $v_g(0,\pm\pi)=v_g(\pm\pi,0)=0$, and others with maximum $v_g(\pm \pi/2,\pm (\mp)\pi/2)$, which explains the non-uniform decay along the different spatial dimensions. Besides, note that due to the finite lifetime of the modes, $\kappa\neq 0$, the localization is exponentially suppressed with the distance of the pumping spot, although not observed clearly due to the saturation of the color scale. When multiples local drives are added to the system, e.g., in a square disposition like in Figs.~\ref{fig:3}(b-c), each local probe funnels excitations mostly into the system in four directions. When the phases of the drives make that the emissions add constructively, all excitations channels are preserved in the steady state, as shown in Fig.~\ref{fig:3}(b). However, if the emissions interfere destructively, the photonic excitations cancel in pairs in all directions, and the light becomes localized between emitters, as illustrated in Fig.~\ref{fig:3}(c). Note, the light localizes in all the region between the pumping spots, as it occurs with emitters~\cite{Gonzalez-Tudela2017b,Gonzalez-Tudela2017a}. The underlying reason is that although the light is preferentially emitted in certain directions, there is light being emitted at all of them, so that light ultimately fills the region in the steady-state. In the extreme case where the lasers are disposed at $(\pm 1,0),(0,\pm 1)$, the localization occurs in the (single) central cavity between the pumping spots. By virtue of the linearity of the steady-state solutions, see Eq.~\eqref{eq:steady1}, summing up several of these solutions one can find any arbitrary pattern in 2D. To illustrate that control, we plot in Fig.~\ref{fig:3}(d) an example of an steady-state localization with a $\Psi$-shape, built out of adding a 4-laser drive for each "pixel" of the letter.

\emph{Conclusions.-} Summing up, we have formalized connections between the spontaneous emission and driven dissipative configuration of structured photonic lattices. These connections bring a simple understanding of the origin of steady-state localization in the driven-dissipative scenario~\cite{Jamadi2021}, as well as the discovery of other phenomena such as directional steady-states or the anti-localization of light. These results opens new avenues in the active control of light, and can pave the way for further experimental works in setups where such structured energy dispersions can be implemented, e.g., photonic-crystal waveguides~\cite{goban13a,goban15a,lodahl15a}, photonic lattices based on coupled microwave resonators~\cite{liu17a,Mirhosseini2018a}, semiconductor microcavities~\cite{Jamadi2021,Boulier2020}, or coupled waveguide arrays~\cite{Christodoulides1988DiscreteWaveguides,Lederer2008DiscreteOptics,Fleischer2003ObservationLattices,Plotnik2011,Weimann2013,rechtsman13a,rechtsman13b,Xu2021}. Even though we illustrate the connection with two particular photonic lattices, it can also provide valuable information to more complex photonic lattices, e.g., with non-trivial topologies~\cite{Bello2019,Gonzalez-Tudela2018,Garcia-Elcano2021} and/or higher-dimensions~\cite{Gonzalez-Tudela2018a}. An interesting outlook of this work is to use this connection for designing arrays of steady-state localized cavities, which can potentially emulate the physics of XY Ising models~\cite{Schneider2017ASimulator,Berloff2017RealizingSimulators,Kalinin2019PolaritonicOscillators,Harrison2020SynchronizationNetworks}.

\begin{acknowledgements}
 AGT acknowledges support from  CSIC Research   Platform   on   Quantum   Technologies   PTI-001  and from  Spanish  project  PGC2018-094792-B-100(MCIU/AEI/FEDER, EU), and from the Proyecto Sinérgico CAM 2020 Y2020/TCS-6545 (NanoQuCo-CM). AGT also acknowledges discussions with the experimental groups of A. Amo, J. Bloch, S. Ravets, whose experimental results of Ref.~\cite{Jamadi2021}, triggered the exploration of these ideas. AGT also thanks Tomas Ramos for a critical reading of the manuscript.
\end{acknowledgements}


\bibliographystyle{apsrev4-2}
\bibliography{references}

\end{document}